# High Expression of CDK1 and NDC80 Predicts Poor Prognosis of Bladder Cancer

**S. Sajedeh Mousavi[1], Mohammad Jalil Zorriehzahra[2*]**

**[1]Department of Veterinary Medicine, Garmsar branch, Islamic azad University, Garmsar, Iran**

**[2]Scientific Information and Communication, Iranian Fisheries Sciences Research Institute(IFSRI), Agriculture Research Education and Extension Organization (AREEO),Tehran, Iran**

*Corresponder author: Mohammad Jalil Zorriehzahra, Scientific Information and Communication, Iranian Fisheries Sciences Research Institute(IFSRI), Agriculture Research Education and Extension Organization (AREEO),Tehran, Iran*

*Tell: +989121075728*

*Email: m.zorriehzahra@areeo.ac.ir*

**Background:** Bladder cancer is the 10$^{th}$ most common cancer worldwide, and its prevalence is increasing, especially in developing countries.

**Objective:** In the present study, we employed gene expression profiles from the GSE163209 data set in the GEO database to identify potential molecular and genetic markers in BC patients.

**Methods:** The data set comprised 217 samples, with 113 stage Ta tumor tissue samples and 104 stage T1 tumor tissue samples. The top 766 genes were chosen. P.value<0.0001 and |logFC|=1 was used to change the cutoff criteria for defining DEGs. Moreover, the MCODE plugin and cytoHubba plugin were employed to produce a module and detect 20 hub genes in these DEGs. We used GO and KEGG pathway enrichment analyses to get a better understanding of these DEGs.

**Results:** The KEGG pathway enrichment results indicated that the top genes were mainly involved: Systemic lupus erythematosus, Alcoholism, and Viral carcinogenesis. SLE activation in the renal glomeruli could explain the connection between this disease's route and bladder cancer, and according to our results and previous researches, heavy alcohol intake can increase the risk of BC in males and particular populations.

**Conclusion:** According to our hub genes, we can consider CDK1 and NDC80 as bladder cancer biomarkers. Not much research has been done on the effect of this gene on bladder cancer.

## 1. Background

Bladder cancer, also known as urological cancer or urinary bladder cancer, is the 10th most common cancer worldwide, and its prevalence is steadily increasing, especially in developing countries. The bladder is a hollow organ located in the lower abdomen that stores urine from the kidneys (via the ureter) before micturition. Urothelial cells, which are specialized transitional epithelial cells that form the urinary bladder and urinary tract, flatten under pressure to accommodate the volume of urine generated. Smooth muscle forms the bladder, which can relax to accommodate larger volumes and contract (voluntarily or reflexively) to eject urine down the urethra and out of the body[2]. The urothelial cells that form the bladder and urinary tract are continuously exposed to potentially mutagenic environmental agents that the kidneys absorb into the urine[9]. Unsurprisingly, these urothelial cells cause 90% of bladder cancer cases, particularly in the developing world, mainly in the bladder but also in the urinary tract on rare occasions. Although localized types of urothelial cancer have a good prognosis, survival rates drop dramatically when the smooth muscle is involved. Squamous cell bladder cancer, which makes up the remaining 10% of cases, is more common in Africa and is thought to be linked to the protozoan infection schistosomiasis[8].

The majority of bladder cancers are linked to exposure to environmental and occupational chemicals, the most common of which is tobacco smoke. The 4-fold gender disparity in bladder cancer incidence can be explained by greater cigarette smoke and occupational exposure in men[4]. The relative risk of bladder cancer associated with tobacco use is only second to lung cancer[13]. In reality, bladder cancer is diagnosed in approximately 80% of adults 65 and older, indicating a disease path that involves decades of exposure or develops decades after exposure[9]. Heritable genetic predispositions have also been linked to bladder cancer in around 7% of cases[1].

## 2. Objectives

In this study, we analyzed the microarray data from GEO (GSE163209) to Identify and evaluate genes that have increased expression in malignant bladder cancer and examine the cellular and molecular pathways in which they are involved.

## 3. Material and methods

### *3.1 Dataset*

The microarray dataset used in this essay was generated by Margaret A Knowles and Hurst CD Based on the GPL17586 Affymetrix Human Transcriptome Array HTA 2.0 platform and is available in NCBI's Gene Expression Omnibus (GEO) repository, with accession number GSE 163209. This study contains 217 bladder tumor samples and obtained from 113 stage Ta and 104 stage T1 bladder tumors.

### *3.2 Preprocessing and Selection of Differentially Expressed Genes (DEGs)*

The gene expression ranges had been converted to a base 2 logarithmic scale. We used R (version 4.0.3) to preprocess the data, and we used various packages such as genefilter (version 1.72.1) to filter genes from throughput experiments, affy (version 1.68.0), Affytools (1.62.0), AffyPLM (1.66.0) to match prob-levels models, and limma (3.46.0) to compare DEGs between Ta and T1 samples. Probe identifiers were also converted into gene symbols using the pd.hta.2.0 annotation package. To be added to the STRING database, the top 766 genes were chosen. P.value<0.0001 and |logFC|=1 was used to change the cutoff criteria for defining DEGs.

*3.3 PPI Undirected Unweighted Network Reconstruction of DEGs*

The online Search Tool for the Retrieval of Interacting Genes (STRING, http://string-db.org) was used to find protein-protein interactions (PPIs) for overlapping DEGs. These interactions were used to build a PPI network of top DEGs using the Cytoscape program (http://www.cytoscape.org/). The Cytoscape MCODE plugin was used to analyze this network. The top 20 upregulated hub genes from the PPI network were chosen using the cytoHubba plugin.

We used Funrich (version 3.1.3) and the David website to consider our top genes, Biological process, Cellular component, Clinical phenotype, Molecular function, and Site of expression.

## 4. Results

*4.1 DEGs gene ontology and KEGG pathway analysis*

To gain further insights into DEGs' biological function, a total of DEGs consisted of 766 DEGs were submitted into DAVID software for GO analysis and KEGG pathway analysis. In biological process (BP) term, the result indicated that the DEGs were mainly enriched in humoral immune response mediated by circulating immunoglobulin, complement activation, chromatin silencing, chromatin silencing at rDNA, DNA packaging, chromosome organization, negative regulation of gene expression(epigenetic), protein activation cascade, DNA replication-dependent nucleosome assembly and humoral immune response. (Figure 1.a). Regarding the cellular component (CC) term, the DEGs were mainly involved in DNA packaging complex, nucleosome, chromosome, protein-DNA complex, nuclear chromosome part, chromosomal region, extracellular region part, chromatin, extracellular exosome, and extracellular vesicle (Figure 1.b). About molecular function (MF), the DEGs were primarily associated with antigen binding, protein heterodimerization activity, serine-type (peptidase activity), serine-type (endopeptidase activity), serine hydrolase activity, protein dimerization activity, histone binding, immunoglobulin receptor binding, endopeptidase activity, and binding (Figure 1.c). The DEGs were mostly linked to Systemic lupus erythematosus, Alcoholism, Viral carcinogenesis, Pentose and glucuronate interconversions, Ascorbate and aldarate metabolism, Steroid hormone biosynthesis, Cell cycle, Retinol metabolism, Porphyrin, and chlorophyll metabolism and Drug metabolism - other enzymes according to the KEGG pathway research. (Figure 1.d).

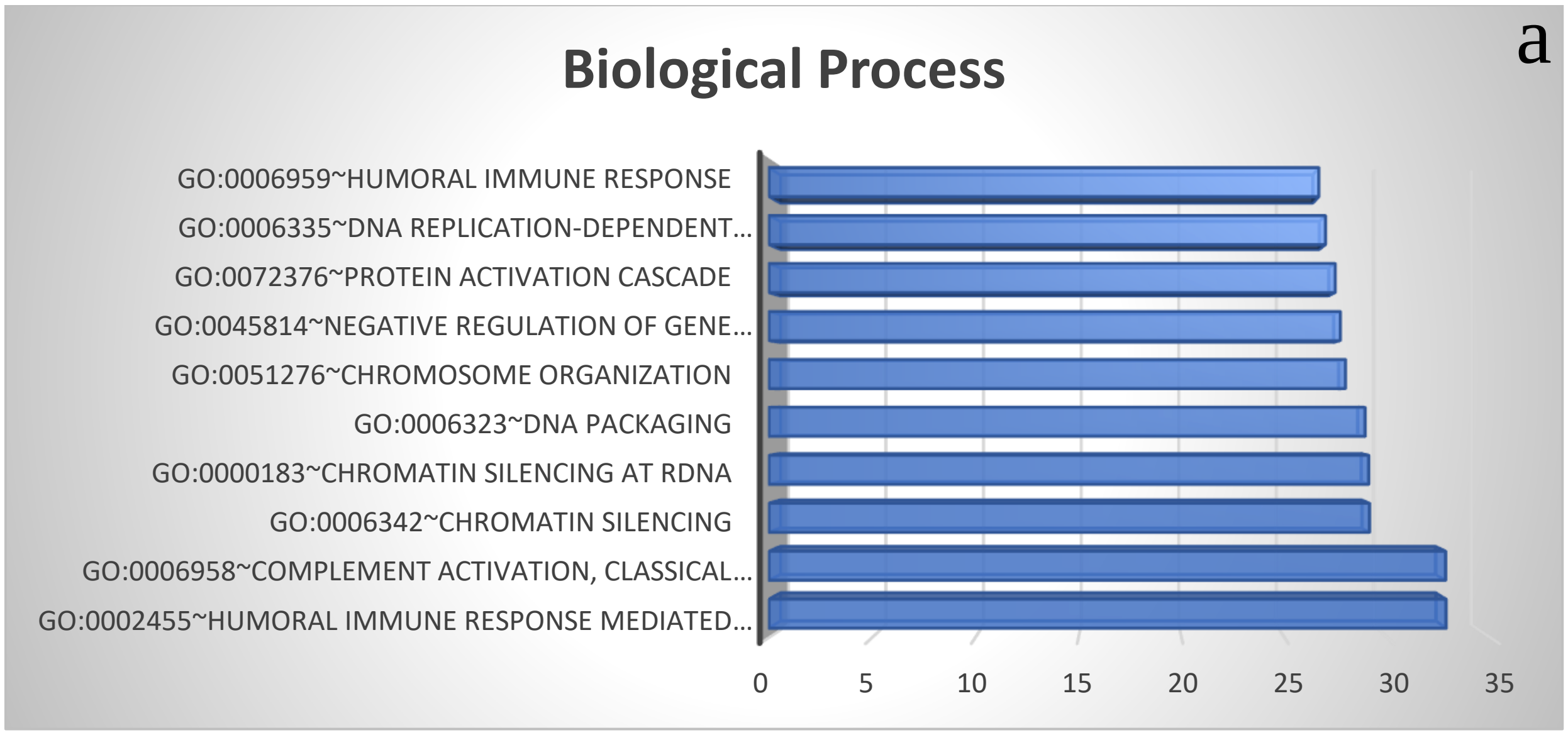

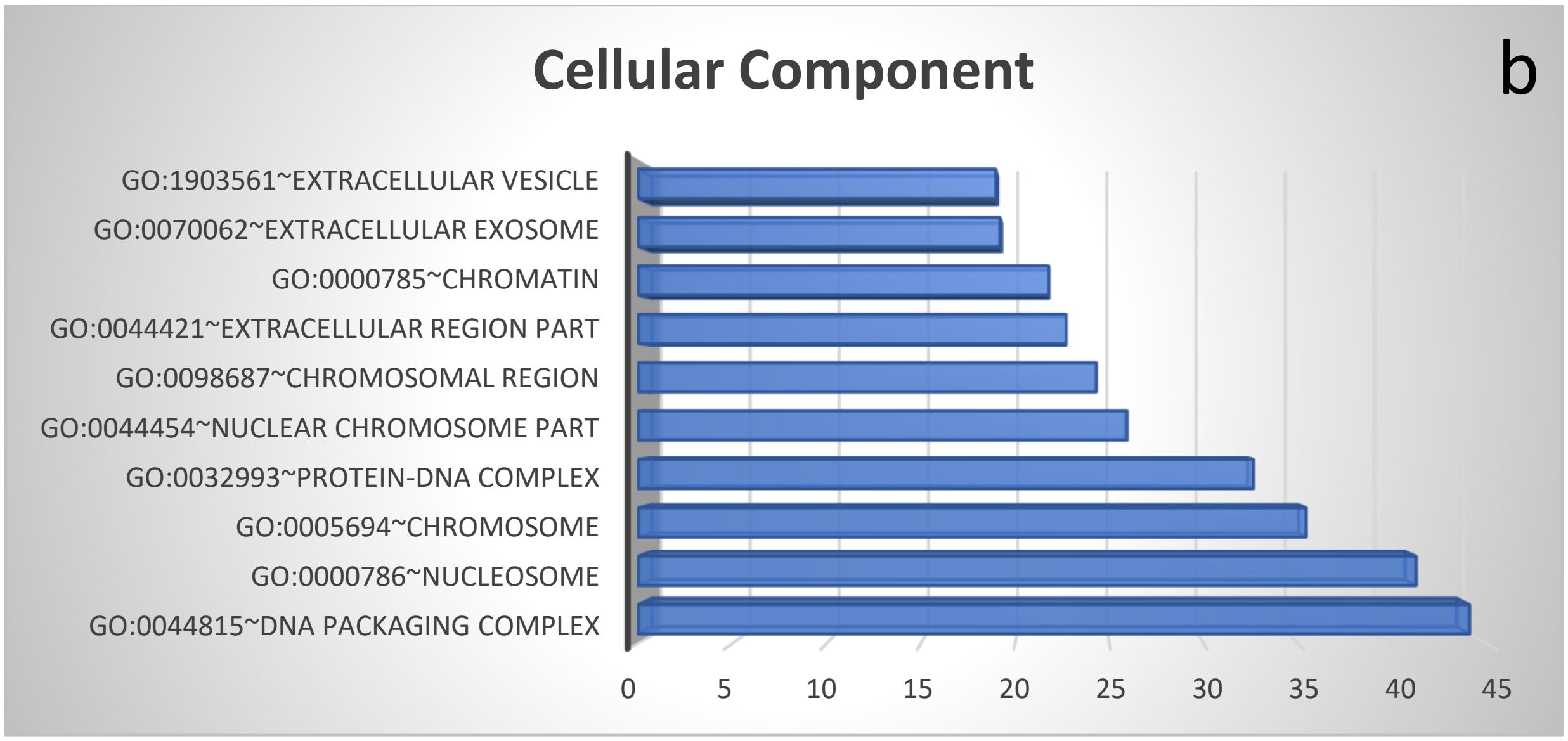

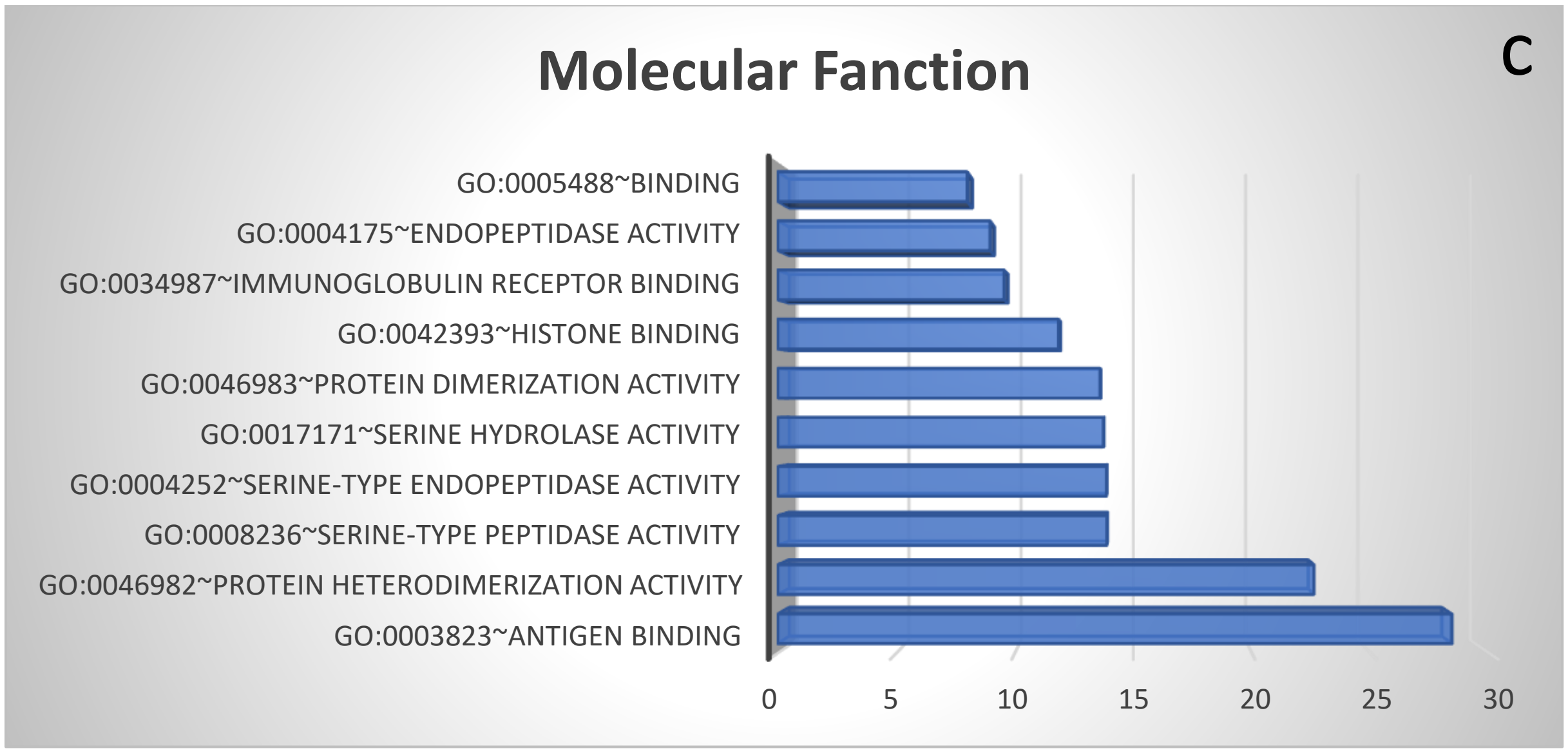

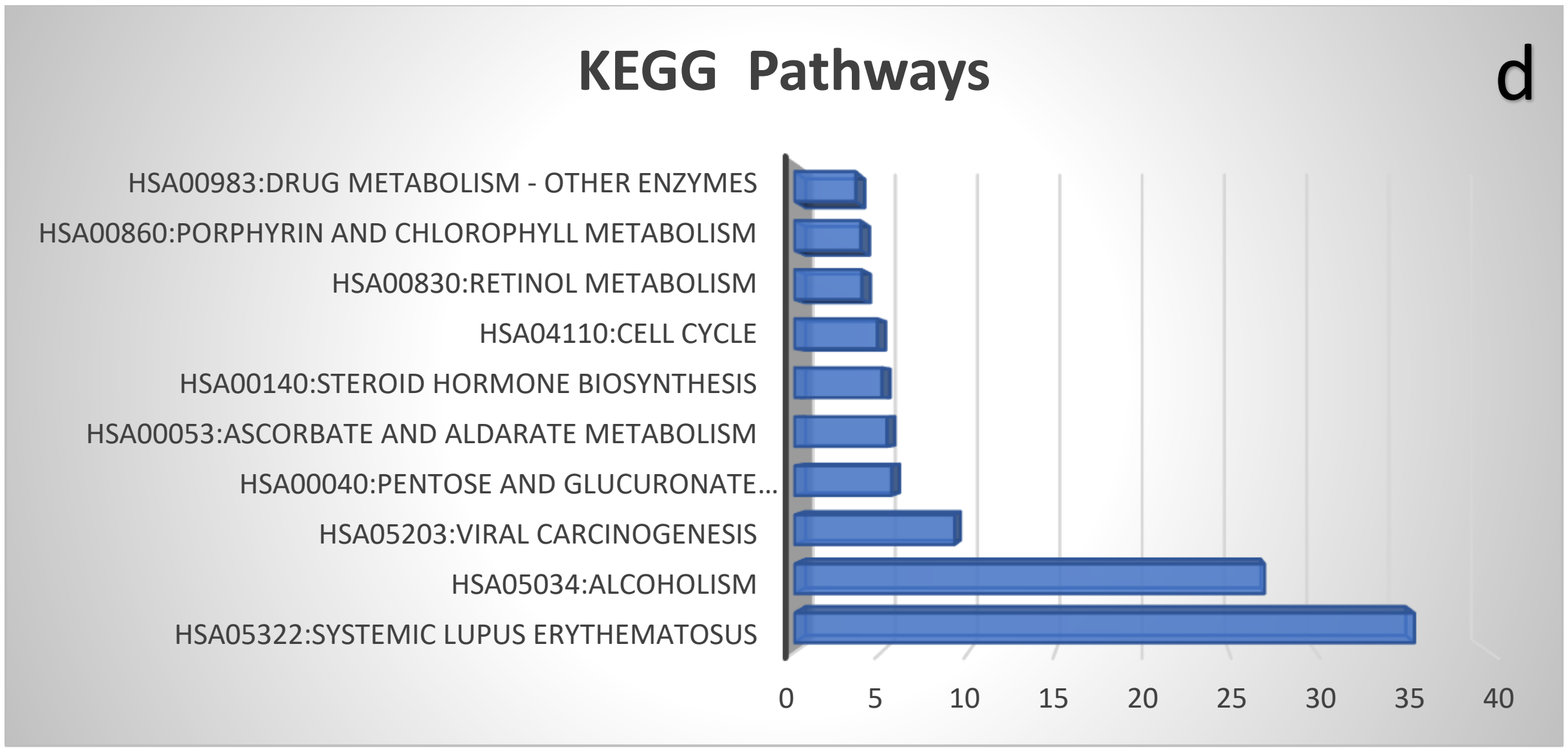

*Figure (1) The involvement degree of our 766 Top genes in different processes: (a)Biological process: Those processes vital for an organism to live and shape its capacities for interacting with its environment. (b)Cellular component: The complex biomolecules and structures of cells and living organisms. (c): Molecular function: the elemental activities of a gene product at the molecular level, such as binding or catalysis. (d)KEGG Pathway: a collection of hand-drawn pathway maps that reflect our understanding of molecular interaction, reaction, and connection networks*

*4.2 PPI network and the top 20 hub genes*

PPI information was acquired from the STRING database (by hiding disconnected nodes in the network), including 554 nodes and 700 edges. The plugin cytoHubba in Cytoscape was used to select the top 20 hub genes, which are in order of scores: CDK1, CCNB1, AURKA, CCNA2, CDC6, BRCA1, TOP2A, BUB1, EZH2, KIF11, NDC80, RAD51, TTK, PLK1, CCNB2, PBK, KIF20A, CDC20, KIF2C and NCAPG

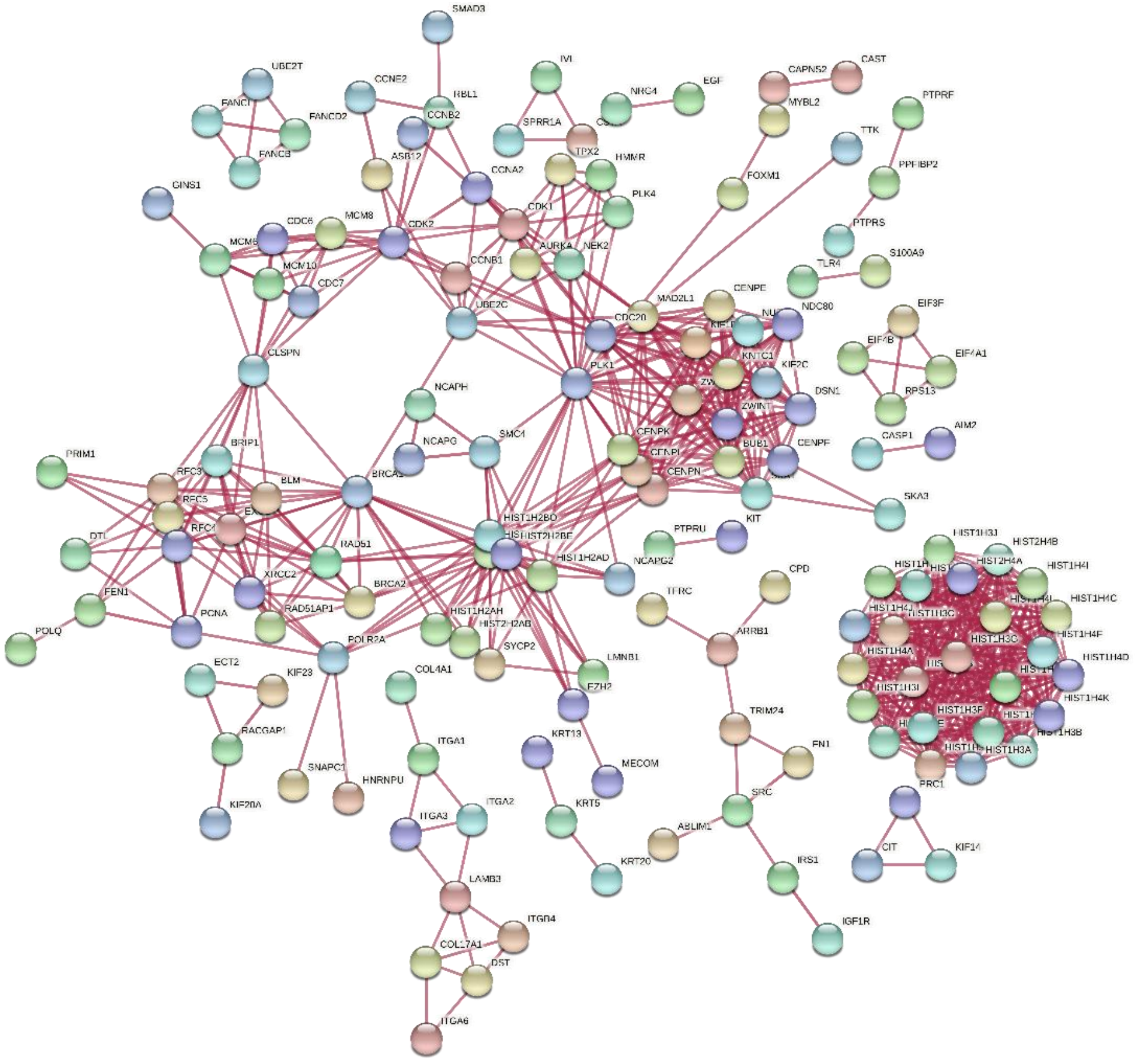

*Figure (2): Protein-protein interactions based on Top genes (by hiding disconnected nodes in the network)*

*4.3 Hub genes analysis*

We used Funrich (version 3.1.3) for various analyses and charts based on our 20 top genes, such as biological process, cellular component, molecular function, heatmap, biological pathways, and interaction diagram. (figure 3)

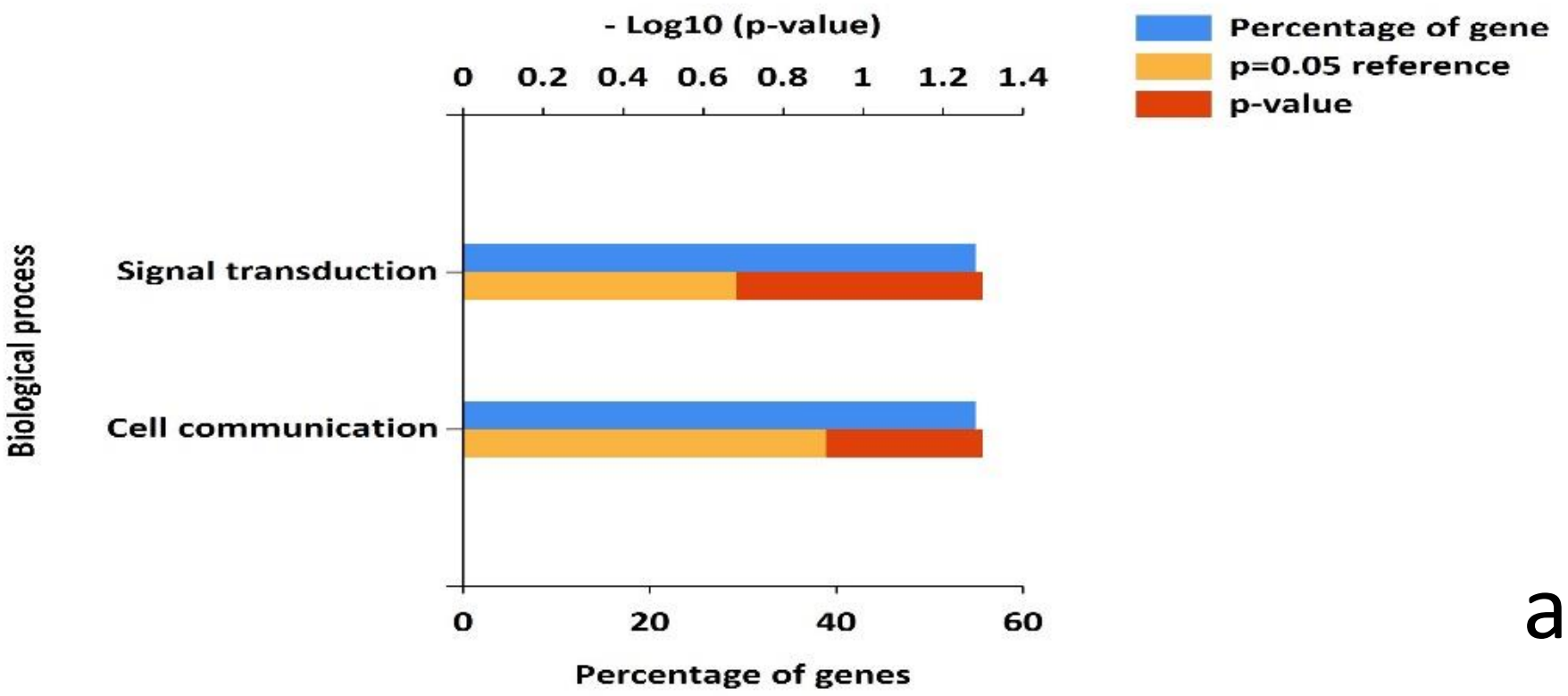

a

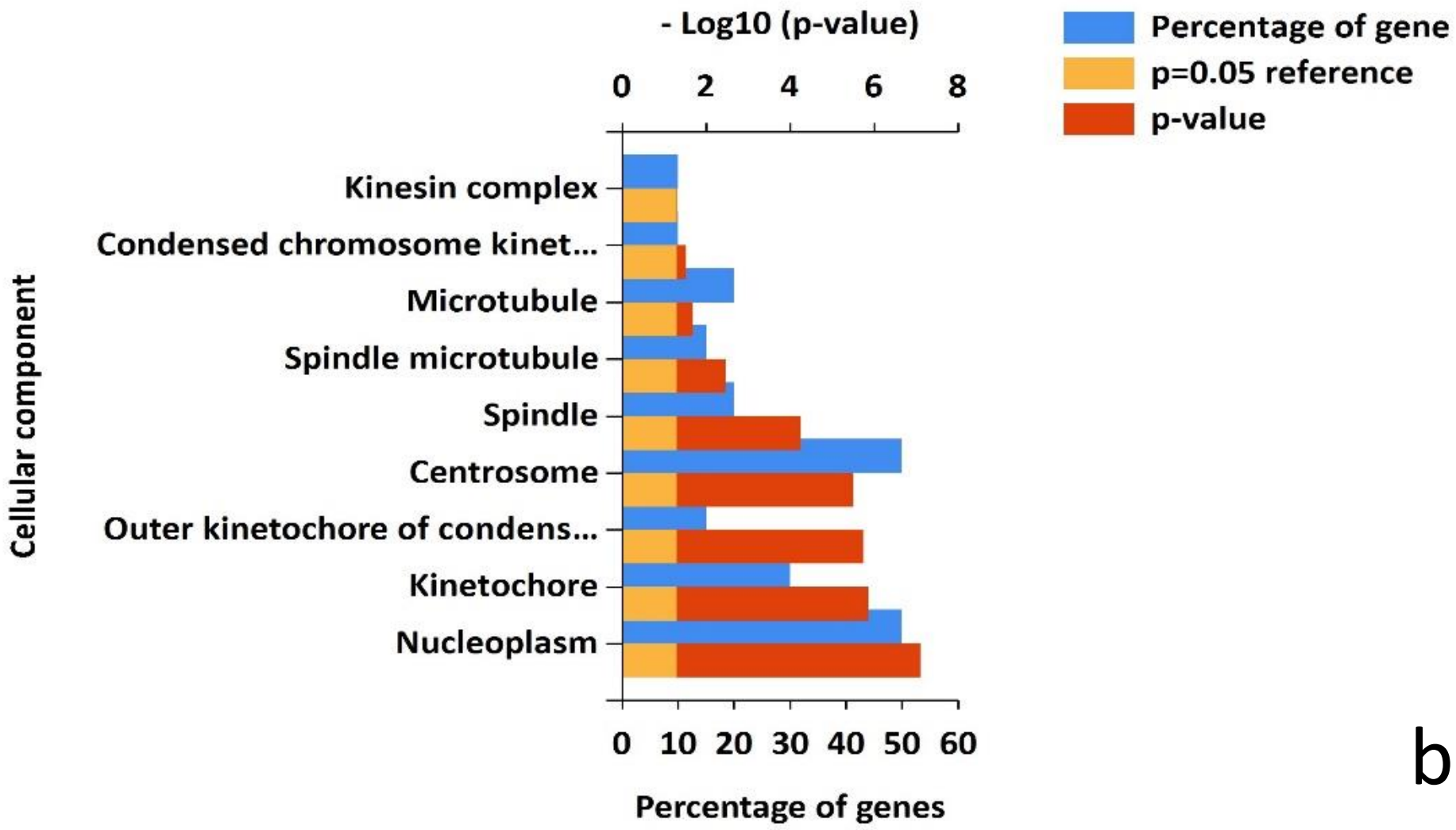

b

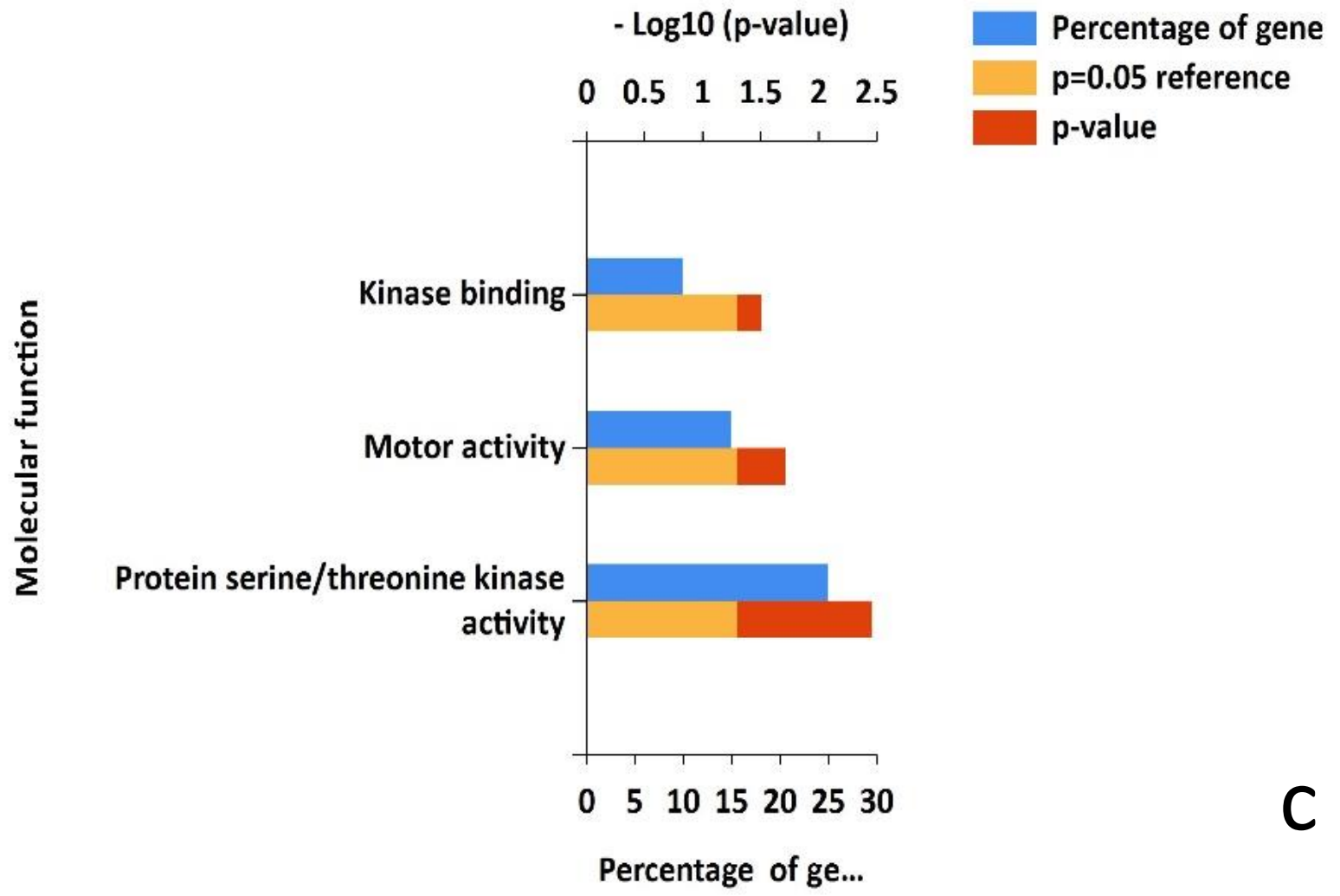
- Log10 (p-value)
0 0.5 1 1.5 2 2.5
Percentage of gene
p=0.05 reference
p-value
Molecular function
Kinase binding
Motor activity
Protein serine/threonine kinase activity
0 5 10 15 20 25 30
Percentage of ge...

c

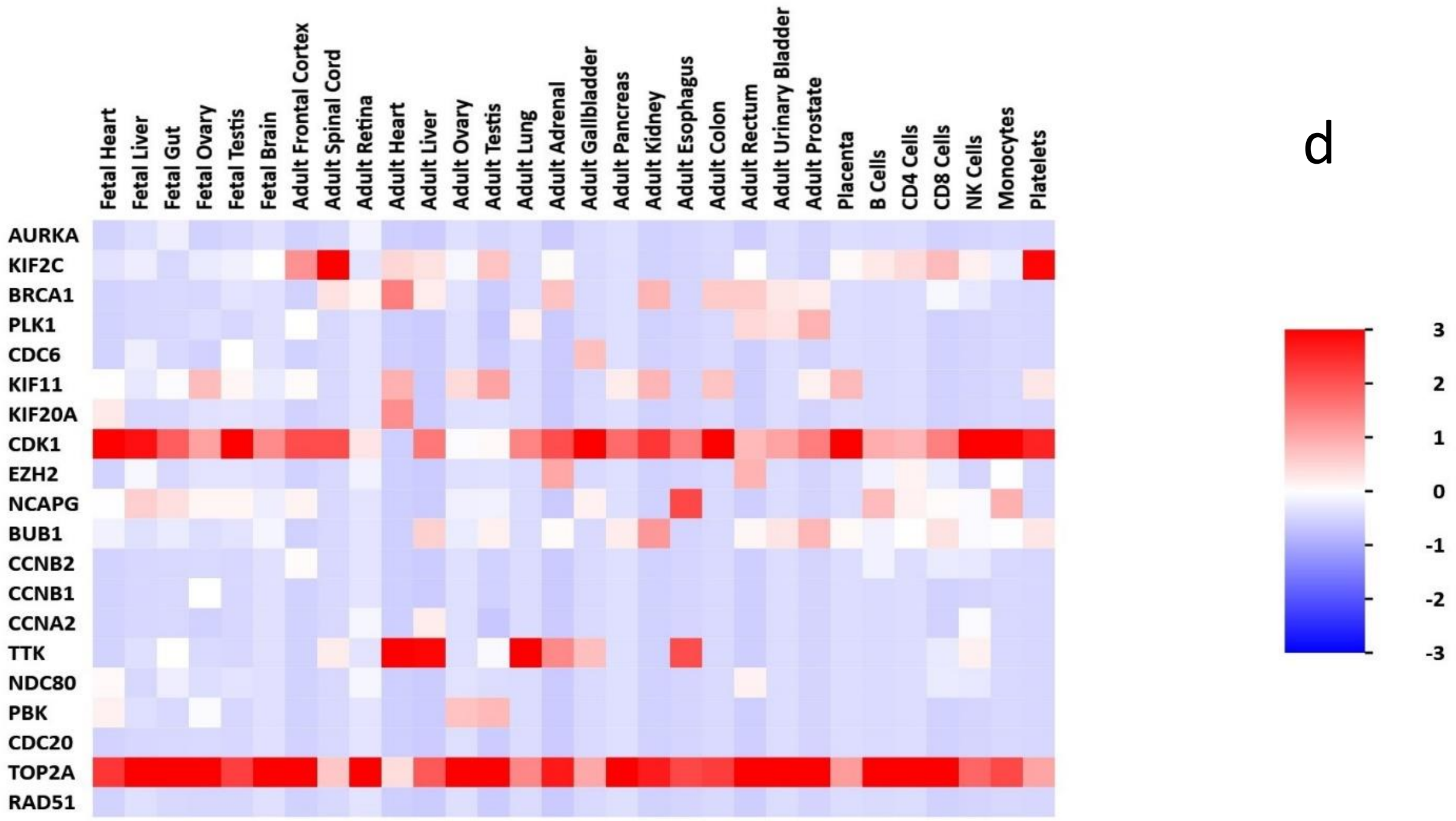

d

e

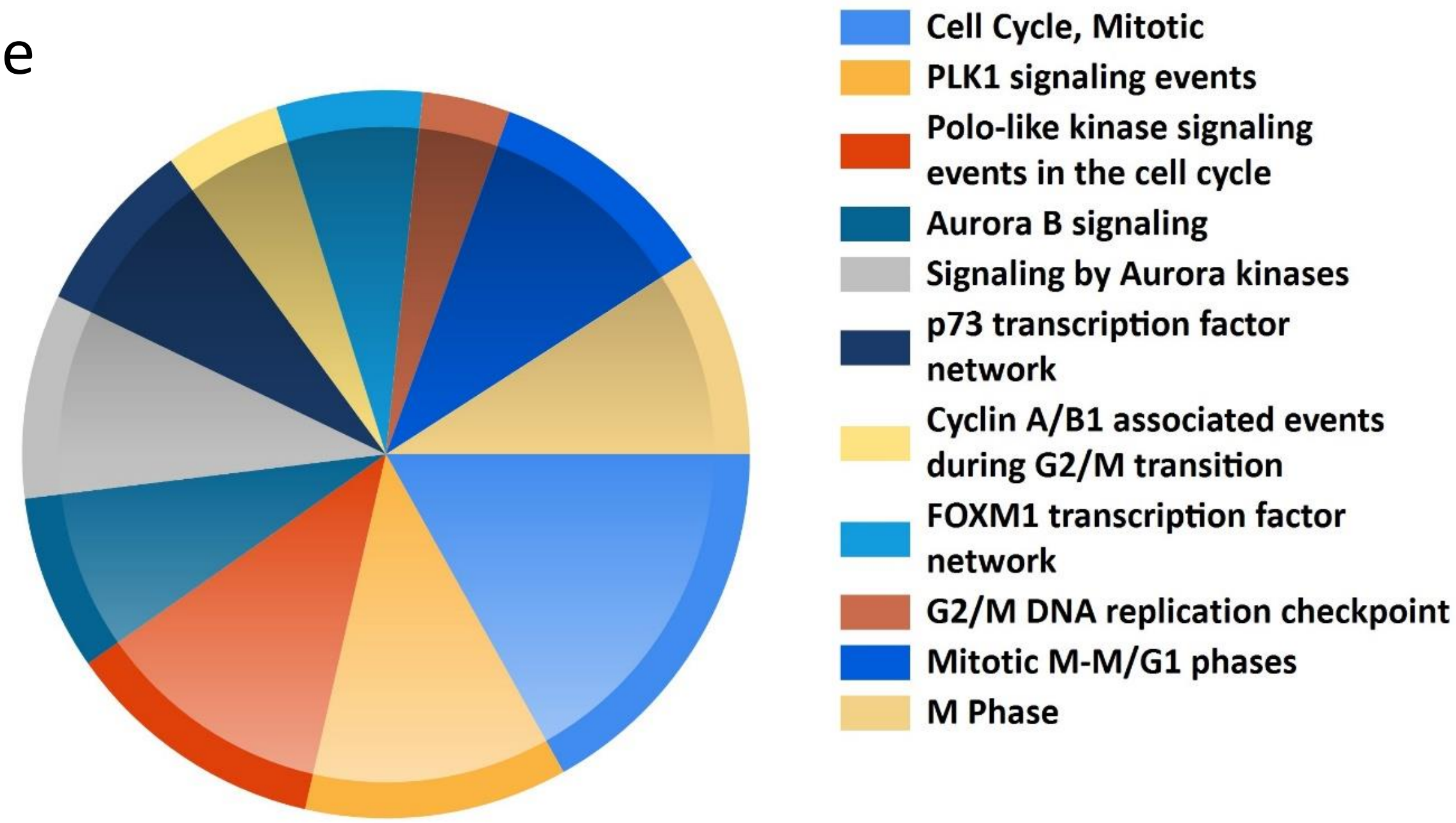
Cell Cycle, Mitotic
PLK1 signaling events
Polo-like kinase signaling events in the cell cycle
Aurora B signaling
Signaling by Aurora kinases
p73 transcription factor network
Cyclin A/B1 associated events during G2/M transition
FOXM1 transcription factor network
G2/M DNA replication checkpoint
Mitotic M-M/G1 phases
M Phase

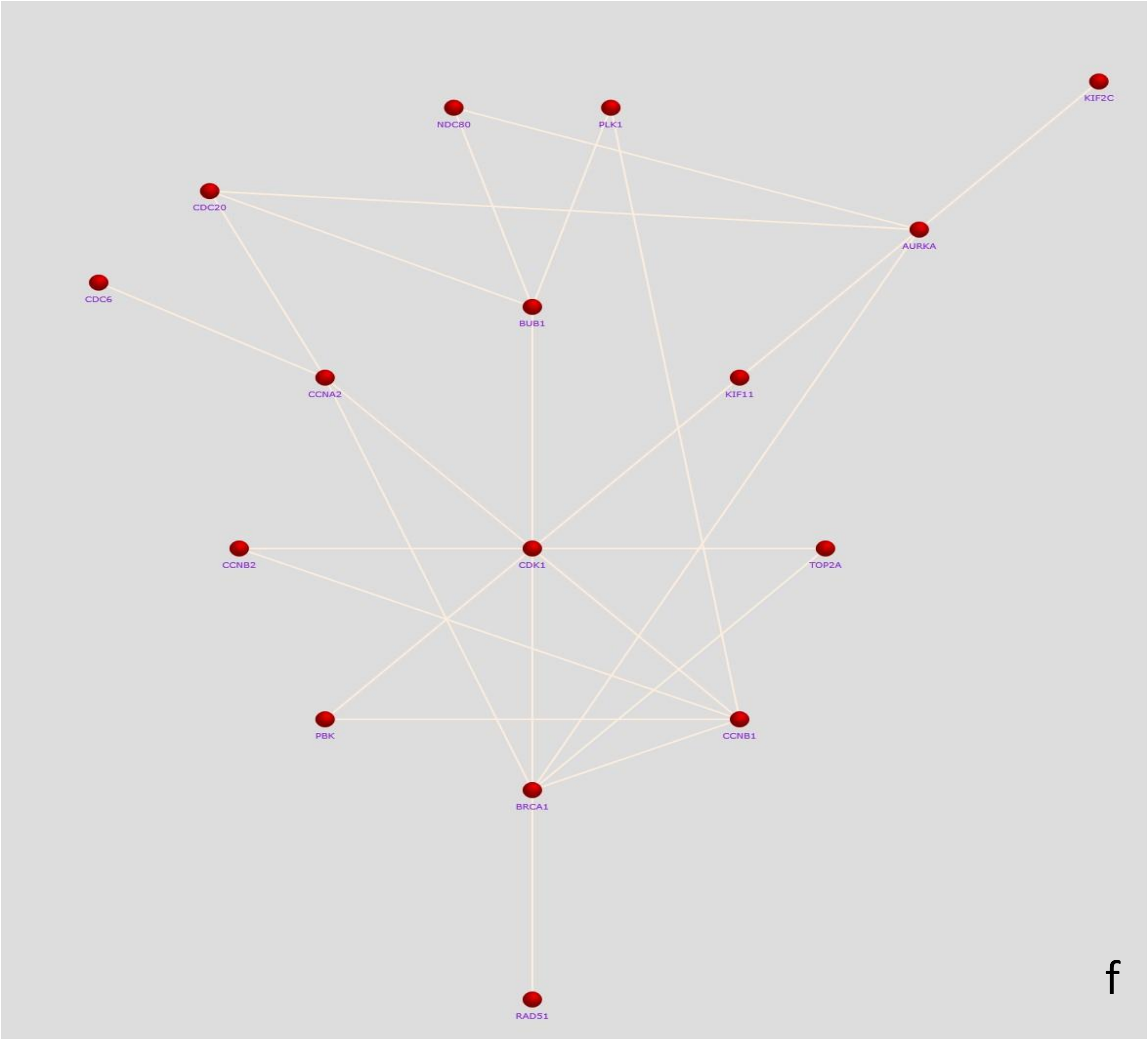

*Figure (3): Analyzing 20 hub genes with Funrich software. (a): Biological process (b)Cellular component (c)Molecular function (d)Heatmap (e)biological pathways (f)Interaction's diagram*

*Table(1): 20 hub genes list with their p.value from our analysis, definitions come from David website*

| Gene Symbol | P.value | Definition |
|---|---|---|
| CDK1 | 3.75E-20 | cyclin dependent kinase 1 |
| CCNB1 | 2.25E-17 | cyclin B1 |
| AURKA | 4.41E-15 | aurora kinase A |
| CCNA2 | 2.44E-26 | cyclin A2 |
| CDC6 | 1.45E-22 | cell division cycle 6 |
| BRCA1 | 2.37E-20 | interacting protein C-terminal helicase 1 |
| TOP2A | 5.36E-31 | topoisomerase (DNA) II alpha |
| BUB1 | 8.72E-19 | BUB1 mitotic checkpoint serine/threonine kinase |
| EZH2 | 8.68-E10 | enhancer of zeste 2 polycomb repressive complex 2 subunit |
| KIF11 | 7.16E-25 | kinesin family member 11 |
| NDC80 | 4.66E-24 | kinetochore complex component |
| RAD51 | 1.06E-17 | recombinase |
| TTK | 5.97E-22 | protein kinase |
| PLK1 | 7.18E.23 | polo-like kinase 1 |
| CCNB2 | 3.37E-19 | cyclin B2 |
| PBK | 2.81E-17 | PDZ binding kinase |
| KIF20A | 2.70E.13 | kinesin family member 20A |
| CDC20 | 6.24E-07 | cell division cycle 20 |
| KIF2C | 1.12E-16 | kinesin family member 2C |
| NCAPG | 2.74E-22 | non-SMC condensin I complex subunit G |

## 5. Discussion

BC is a genitourinary tumor that arises from the epithelial lining of the urinary bladder and is one of the most common. Pathological analyses, including the clinical and tumor grades, are currently the critical determinants for risk assessment and therapeutic decision-making in BC patients. [3]However, none of the conventional histopathological parameters have adequate sensitivity and specificity for detecting, controlling, and predicting prognosis in BC patients. [6]Many studies have attempted to identify possible molecular markers for early detection, early diagnosis, and the development of effective treatments due to these limitations.

In the present study, we employed gene expression profiles from the GSE163209 data set in the GEO database to identify potential molecular and genetic markers in BC patients. The data set comprised 217 samples, with 113 stage Ta tumor tissue samples and 104 tumor tissue samples. The top 766 genes were chosen. P.value<0.0001 and |logFC|=1 was used to change the cutoff criteria for defining DEGs. Moreover, the MCODE plugin and cytoHubba plugin were employed to produce a module and detect 20 hub genes in these DEGs. We used GO and KEGG pathway enrichment analyses to get a better understanding of these DEGs.

The KEGG pathway enrichment results indicated that the top genes were mainly involved: Systemic lupus erythematosus, Alcoholism, Viral carcinogenesis, Pentose and glucuronate interconversions, Ascorbate, and aldarate metabolism, Steroid hormone biosynthesis, Cell cycle, Retinol metabolism, Porphyrin, and chlorophyll metabolism, and Drug metabolism - other enzymes.

The outcomes show that Systemic lupus erythematosus is one of the most critical pathways in bladder cancer. SLE is an autoimmune disorder marked by the formation of IgG autoantibodies against self-antigens such as DNA, nuclear proteins, and various cytoplasmic components and a variety of clinical symptoms. Inflammation, vasculitis, immune complex deposition, and vasculopathy are the most common pathological findings in patients with SLE. Immune complexes containing autoantibody and self-antigen are deposited in the renal glomeruli, where they activate complement or Fc[9]R- mediated neutrophil and macrophage activation, resulting in a systemic inflammatory response. The formation of the attack complex on the membrane (C5b-9) or the generation of the anaphylatoxin and cell activator C5a results in injury when complement (C5) is activated. The release of oxidants and proteases by neutrophils and macrophages causes tissue injury (genome website). SLE activation in the renal glomeruli could explain the connection between this disease's route and bladder cancer.

The second one in our Kegg list is Alcoholism. Alcoholism, also known as alcohol dependency (ethanol), is a persistent relapsing condition that progresses and has significant adverse health consequences. Dopaminergic ventral tegmental region (VTA) projections to the nucleus accumbens (NAc) have been described as one of the critical mediators of alcohol's rewarding effects. Acute alcohol exposure induces dopamine release into the NAc, which activates D1 receptors, stimulating PKA signaling and CREB-mediated gene expression, while chronic alcohol exposure causes an adaptive downregulation of this pathway, especially CREB function. Reduced CREB function in the NAc may encourage illegal drugs to increase reward and regulate positive affective states in addiction. PKA signaling influences NMDA receptor activity and can play a role in neuroadaptation following chronic alcohol exposure (genome website). According to a 2019 study by Mihai Dorin Vartolomei, heavy alcohol intake can increase the risk of BC in males and particular populations. In addition to smoking cessation programs, alcohol cessation services should be used in BC prevention efforts because they can raise the risk of BC while also causing various other diseases. [12]

From the Kegg chart, it seems Viral carcinogenesis could be involved in BC. Viruses and the development of human cancers have a close connection. At least six human viruses, Epstein-Barr virus (EBV), hepatitis B virus (HBV), hepatitis C virus (HCV), human papillomavirus (HPV), human T-cell lymphotropic virus (HTLV-1), and Kaposi's associated sarcoma virus (KSHV), are now known to lead to 10-15 percent of cancers worldwide. These tumor viruses facilitate abnormal cell proliferation by modulating cellular cell-signaling pathways and evading cellular defense mechanisms such as blocking apoptosis through the expression of numerous potent oncoproteins. Oncoproteins from human tumor viruses may also interfere with pathways that keep the host cell's genome intact. Viruses that encode these behaviors can play a role in the initiation and progression of human cancers. (genome website)

For example, A study by Babak Javanmard in 2019 conclude that HPV infection has been linked to more advanced stages and grades of bladder cancer[5]. Also, research by Yigang Zeng in 2020 shows that BKPyV (BK polyomavirus) infection promotes the proliferation, invasion, and migration of bladder cancer. They verified the role of the β-catenin signaling pathway and the Epithelial-Mesenchymal Transition effect in BKPyV-infected bladder cancer. These results provide meaningful information towards the diagnosis and treatment of clinical bladder cancer. [14]

The first gene with the best score in the hub genes list is CDK1. Not much research has been done on the effect of this gene on bladder cancer; however, one study in 2018 shows lncRNA PVT1–miR-31–CDK1 pathway was involved in the development of bladder cancer, but the number of samples was not enough (35 samples) and our analysis can confirm their results[11]. The activity of several activators (cyclins) and inhibitors regulates the activity of cyclin-dependent kinases (CDKs), which control the mammalian cell cycle (Ink4, and Cip and Kip inhibitors). Cell cycle CDKs are deregulated in cancer cells due to genetic or epigenetic changes in CDKs, their regulators, or upstream mitogenic pathways.[7]

Another vital gene in our list is NDC80 which in just one research in 2017 by Zhe Zhang introduce as a BC biomarker. Cause this gene has a great P.value in our analysis results (4.66E-24), it may be an excellent choice to consider as a BC biomarker. (comprehensive) Research by Ngang Heok Tang and Tkashi Toda in 2015 mentioned that they hypothesise that overproduced Ndc80/Hec1 may sequestrate its binding partners through the Ndc80 internal loop domain. This absorption will result in altered mitotic progression and defective chromosome segregation, leading to aneuploidy.

On the other hand, in response to these defects, the transcription program of NdC80 and its interacting partners may also be altered due to different cell cycle profiles and cellular compensatory mechanisms. This alteration may further promote the growth of aneuploid cells. Identification of the Ndc80 internal loop as a protein-protein interaction motif has shed light on our understanding of Ndc80 overexpression and cancer formation. [10]

## 6. Conclusion

In conclusion, Systemic lupus erythematosus, Alcoholism, and Viral carcinogenesis could be involved in poor prognosis of bladder cancer, and overexpression of CDK1 and NDC80 could be considered as BC biomarkers.

## Acknowledgments

All the authors acknowledge the support from their respective institutions and universities.